\documentclass[dvips]{acta}
\usepackage{supertabular,lscape,epsfig}
\usepackage{amssymb}
\usepackage{amsmath}
\usepackage{graphicx}
\usepackage{txfonts}
\usepackage[T1]{fontenc}

\SetPages{0}{0}

\SetVol{58}{2008}

\begin{document}

\begin{Titlepage}
\Title{An X-Shooter view of the symbiotic star [JD2002]\,11\thanks{Based on observations collected at the European Organisation for Astronomical Research in the Southern Hemisphere, Chile (291.D-5044(A)).}}

\Author{Hajduk$^1$, M., Gromadzki$^{2,3}$, M., Miko{\l}ajewska$^1$, J., Miszalski$^{4,5}$, B., and Soszy\'{n}ski$^6$, I.}{$^1$Nicolaus Copernicus Astronomical Center, ul. Bartycka 18, 00-716, Warsaw, Poland\\
e-mail:cinek@ncac.torun.pl,mikolaj@ncac.torun.pl\\
$^2$Millennium Institute of Astrophysics, Av. Vicua Mackenna 4860, 782-0436, Macul, Santiago, Chile\\
$^3$Universidad de Valpara{\'i}so, Av. Gran Breta\~{n}a 1111, Playa Ancha, Casilla, 5030, Chile\\
e-mail:mariusz.gromadzki@uv.cl\\
$^4$South African Astronomical Observatory, PO Box 9, Observatory, 7935 Cape Town, South Africa\\
$^5$Southern African Large Telescope Foundation, PO Box 9, Observatory, 7935 Cape Town, South Africa\\
e-mail:brent@saao.ac.za\\
$^6$Warsaw University Observatory, Al. Ujazdowskie 4, 00-478 Warszawa, Poland\\
e-mail:soszynsk@astrouw.edu.pl}

\Received{Month Day, Year}
\end{Titlepage}

\Abstract{We aimed to verify the nature and derive the basic parameters of
the symbiotic star candidate [JD2002]\,11. For this purpose, we obtained
and analysed an X-Shooter spectrum of [JD2002]\,11. We also used optical
and infrared photometry available for the object. Emission-line diagnostic
ratios are characteristic of a dusty type symbiotic star and reveal a
two-component nebula (low- and high-density). The spectral energy distribution
is well fitted with a two-component blackbody spectrum with the respective
temperatures of 1150\,K and 600\,K. The total luminosity of $\rm
3000\,L_{\odot}$ is consistent with the expected luminosity of a typical Mira
star, embedded in an optically thick dust shell. We conclude that
[JD2002]\,11 is the ninth symbiotic star in total and only the second dusty type
symbiotic star discovered in the Small Magellanic Cloud.}
{binaries: symbiotic -- planetary nebulae: general -- stars: individual: [JD2002]\,11 -- Magellanic Clouds}

\section{Introduction}

Symbiotic stars are the products of binary evolution of low mass stars. These
systems comprise of a giant star and a compact (usually white dwarf) companion.
The giant star looses mass via stellar wind or the Roche lobe, part of which is
accreted by the hot component. 

Most of symbiotic stars belong to the so-called S-type (stellar), showing the
presence of stellar photospheres of normal red giants. In D-type (dusty)
symbiotics the cool component is a Mira star enshrouded in a warm dust shell 
(Miko{\l}ajewska 2013).

Population of extragalactic symbiotic systems is important for our
understanding of the symbiotic phenomenon, since the sample includes objects
with known distances and with diverse chemical compositions. Currently, there
are eight symbiotic stars known in the SMC (Belczy{\'n}ski et al., 2000;
Oliveira et al., 2013; Miszalski et al., 2014). The only one D-type system
in this galaxy was found by Oliveira et al. (2013). [JD2002]\,11 is a D-type
system candidate proposed by Hajduk et al. (2014).

In this paper we present spectroscopic observations of [JD2002]\,11 obtained
using X-Shooter mounted on the Very Large Telescope UT2. The spectrograph
provides a stunning wavelength coverage, which we used to verify the
symbiotic nature of the object.

\section{Observations and data reduction}

\subsection{The X-Shooter spectrum}

Two exposures of [JD2002]\,11 were performed with X-Shooter in the UVB, VIS, and
NIR arm in the nodding mode on October 13, 2013. An artificial strip was present
in the second VIS frame and the observations were interrupted and repeated on
the following night. However the rest of the data was not corrupted. The target
was observed at the airmass of 1.6. Feige\,110 was used as the flux standard,
observed at the airmass of 1.1.

Three spectra of [JD2002]\,11 in each arm were obtained on October 14, 2013 at
the airmass of 1.6. Two exposures were performed in the nodding mode and one in
the stare mode. The spectrum of GD\,71, obtained at the airmass of 1.3, was used
for flux calibration.

Seeing was slightly better during the first night (0.8 arcsec compared to 1.0
arcsec). Thin clouds were present on the sky on both nights.

Observations in three arms started simultaneously and took 865 s in the UVB arm, 930 s in the VIS arm, and 900 s in the NIR arm.

The observations of [JD2002]\,11 were performed with the $1.0 \times 11$ arcsec
slit for the UVB arm and $0.9 \times 11$ arcsec for the VIS and NIR arms.
Resolution ranges from 5100 in the UVB and NIR arms to 8800 in the VIS arm. Flux
standards were observed using the $5 \times 11$ arcsec slit for each arm. 

The data were reduced with the X-Shooter pipeline version 2.5.2 provided by ESO.
This included the wavelength calibration, spectrum subtraction, and flux
calibration. The spectra were reduced with the xsh\_scired\_slit\_stare (UVB and
VIS arms) and xsh\_scired\_slit\_nod (NIR arm) recipes. The latter recipe
allowed for optimal subtraction of the sky lines. The correction for the
telluric lines was performed using the telluric standard observation using IRAF
package\footnote{IRAF is distributed by the National Optical Astronomy
Observatory, which is operated by the Association of Universities for Research
in Astronomy (AURA) under cooperative agreement with the National Science
Foundation.}. 

The spectra taken at two nights were examined and compared. No significant
differences were found. Thus we averaged all the spectra in each arm to achieve
better signal to noise ratio. The VIS and NIR fluxes were increased by 6\%
relative to UVB fluxes to compensate for the smaller slit width. The relative
flux calibration uncertainty of 5\% was adopted from comparison of the response functions determined for the two nights. The line fluxes are presented
in Table~1.

\begin{table}
\caption{Emission line fluxes in the X-Shooter spectra.}
\label{lines}
\centering
{\tiny
\begin{tabular}{c c c c c c}
\hline\hline
$\rm \lambda_{obs}$ [\AA]& ident & $\rm \lambda_{lab}$ [\AA]& flux & flux error &der. flux\\
\hline
3688.764&H\,{\sc i}	&3686.83&	1.59	&	0.67&	1.81	\\
3693.654&H\,{\sc i}	&3691.56&	1.39	&	0.56&	1.58	\\
3699.009&H\,{\sc i}	&3697.15&	2.38	&	0.77&	2.71	\\
3705.942&H\,{\sc i}	&3703.86&	2.79	&	0.82&	3.17	\\
3713.736&H\,{\sc i}	&3711.97&	1.63	&	0.67&	1.85	\\
3723.657&H\,{\sc i}	&3721.94&	2.67	&	0.31&	3.02	\\
3728.053&[O\,{\sc ii}]	&3726.03&	13.85	&	0.75&	15.71	\\
3730.786&[O\,{\sc ii}]	&3728.82&	14.98	&	0.80&	16.98	\\
3736.209&H\,{\sc i}	&3734.37&	2.97	&	0.41&	3.36	\\
3751.918&H\,{\sc i}	&3750.15&	3.25	&	0.39&	3.68	\\
3772.551&H\,{\sc i}	&3770.63&	4.18	&	0.41&	4.72	\\
3799.820&H\,{\sc i}	&3797.9	&	5.23	&	0.42&	5.88	\\
3821.492&He\,{\sc i}	&3819.62&	1.11	&	0.25&	1.24	\\
3837.252&H\,{\sc i}	&3835.39&	7.62	&	0.51&	8.53	\\
3841.347&[Fe\,{\sc v}]	&3839.27&	3.08	&	0.41&	3.44	\\
3870.663&[Ne\,{\sc iii}]&3868.75&	45.12	&	2.28&	50.35	\\
3890.906&H\,{\sc i}+ He	&3889.05&	17.69	&	0.95&	19.69	\\
3893.039&[Fe\,{\sc v}]	&3891.28&	5.75	&	0.56&	6.40	\\
3897.359&[Fe\,{\sc v}]	&3895.22&	2.54	&	0.46&	2.83	\\
3969.395&[Ne\,{\sc iii}]&3967.46&	15.59	&	0.87&	17.21	\\
3971.999&H\,{\sc i}	&3970.07&	17.38	&	0.94&	19.18	\\
4028.210&He\,{\sc i}	&4026.19&	2.32	&	0.37&	2.54	\\
4070.753&[S\,{\sc ii}]	&4068.6	&	1.57	&	0.34&	1.71	\\
4103.718&H\,{\sc i}	&4101.74&	27.48	&	1.40&	29.90	\\
4145.833&He\,{\sc i}	&4143.76&	0.93	&	0.28&	1.00	\\
4342.570&H\,{\sc i}	&4340.74&	49.67	&	2.50&	52.71	\\
4365.224&[O\,{\sc iii}]	&4363.23&	33.82	&	1.72&	35.81	\\
4473.663&He\,{\sc i}	&4471.5	&	5.00	&	0.35&	5.23	\\
4687.721&He\,{\sc ii}	&4685.58&	14.38	&	0.77&	14.68	\\
4703.835&[Fe\,{\sc iii}]&4701.62&	0.79	&	0.21&	0.80	\\
4715.942&He\,{\sc i}	&4713.17&	3.56	&	0.38&	3.63	\\
4726.588&[Ne\,{\sc iv}]	&4724.15&	2.97	&	0.39&	3.02	\\
4734.299&[Fe\,{\sc iii}]&4733.93&	1.49	&	0.41&	1.51	\\
4742.592&[Ar\,{\sc iv}]	&4740.17&	1.43	&	0.39&	1.45	\\
4863.669&H\,{\sc i}	&4861.33&	100.00	&	5.01&	100.00	\\
4924.261&He\,{\sc i}	&4921.93&	0.99	&	0.23&	0.99	\\
4961.441&[O\,{\sc iii}]	&4958.91&	79.71	&	3.99&	78.79	\\
4969.196&[Fe\,{\sc vi}]	&4967.32&	0.94	&	0.28&	0.93	\\
4974.653&[Fe\,{\sc vi}]	&4972.5	&	0.96	&	0.29&	0.95	\\
5009.392&[O\,{\sc iii}]	&5006.84&	237.39	&	11.87&	233.31	\\
5018.138&He\,{\sc i}	&5015.68&	1.72	&	0.20&	1.69	\\
5147.379&[Fe\,{\sc vi}]	&5146.8	&	1.86	&	0.29&	1.80	\\
5178.143&[Fe\,{\sc vi}]	&5176.43&	2.37	&	0.26&	2.28	\\
5311.436&[Ca\,{\sc v}]	&5309.11&	0.78	&	0.19&	0.74	\\
5413.778&He\,{\sc ii}	&5411.52&	0.97	&	0.24&	0.91	\\
5878.494&He\,{\sc i}	&5875.66&	27.73	&	1.47&	25.05 	\\
6303.668&[O\,{\sc i}]	&6300.34&	4.25	&	0.33&	3.72  	\\
6315.377&[S\,{\sc iii}]	&6312.1	&	1.14	&	0.25&	1.00  	\\
6367.250&[O\,{\sc i}]	&6363.78&	1.54	&	0.29&	1.34  	\\
6551.608&[N\,{\sc ii}]	&6548.1	&	3.19	&	0.23&	2.75  	\\
6566.161&H\,{\sc i}	&6562.77&	464.46	&	23.22&	399.41	\\
6586.981&[N\,{\sc ii}]	&6583.5	&	9.06	&	0.49&	7.78  	\\
6681.424&He\,{\sc i}	&6678.16&	6.01	&	0.37&	5.13 	\\
6719.922&[S\,{\sc ii}]	&6716.44&	2.34	&	0.23&	1.99 	\\
6734.357&[S\,{\sc ii}]	&6730.82&	2.17	&	0.22&	1.85 	\\
7068.701&He\,{\sc i}	&7065.25&	18.67	&	0.95&	15.57	\\
7139.464&[Ar\,{\sc iii}]&7135.8	&	3.58	&	0.23&	2.97 	\\
7174.147&[Ar\,{\sc iv}]	&7170.86&	0.39	&	0.10&	0.32 	\\
7285.132&He\,{\sc i}	&7281.35&	1.19	&	0.14&	0.98 	\\
7323.767&[O\,{\sc ii}]	&7319.45&	4.94	&	0.31&	4.06 	\\
7334.073&[O\,{\sc ii}]	&7330.2	&	4.39	&	0.29&	3.61 	\\
7755.305&[Ar\,{\sc iii}]&7751.06&	1.25	&	0.18&	1.01 	\\
8442.368&H\,{\sc i}	&8438	&	0.36	&	0.14&	0.28 	\\
8506.963&H\,{\sc i}	&8502.49&	0.99	&	0.19&	0.77 	\\
8549.716&H\,{\sc i}	&8545.38&	1.16	&	0.19&	0.90 	\\
8602.632&H\,{\sc i}	&8598.39&	1.78	&	0.22&	1.38 	\\
8669.150&H\,{\sc i}	&8665.02&	1.69	&	0.22&	1.31 	\\
8754.766&H\,{\sc i}	&8750.48&	1.64	&	0.19&	1.27 	\\
9019.435&H\,{\sc i}	&9015	&	4.20	&	0.29&	3.22	\\
9073.730&[S\,{\sc iii}]	&9068.6	&	3.58	&	0.24&	2.74	\\
9233.570&H\,{\sc i}	&9229.01&	4.64	&	0.34&	3.53	\\
9536.008&[S\,{\sc iii}]	&9530.6	&	7.86	&	0.49&	5.93	\\
9550.665&H\,{\sc i}	&9545.97&	8.71	&	0.52&	6.57	\\
10054.47&H\,{\sc i}	&10049.37&	12.23	&	0.96&	9.11	\\
10127.69&He\,{\sc ii}	&10123.61&	9.59	&	0.96&	7.13	\\
10835.97&He\,{\sc i}	&10829.894&	182.39	&	9.13&	133.59	\\
10942.99&H\,{\sc i}	&10938.095&	12.69	&	0.81&	9.27  	\\
12823.65&H\,{\sc i}	&12818.08&	29.85	&	1.61&	21.21 	\\
\hline
\end{tabular}
}
\end{table}

\subsection{Other observations}

The spectrum of the object was also obtained with the Southern African Large Telescope (SALT) by Hajduk et al. (2014). The wavelength range of the spectrum is from 4332\,\AA\ to 7415\,\AA. The seeing during the observation was about 3 arcsec. [JD2002]\,11 is spatially blended in this spectrum with an early-type background star, which gives rise to a continuum. Hydrogen and helium stellar lines are blended with the emissions, which may affect the measurements. Resolution of the spectrum was about 1000. Sensitivity curve from another night was used for flux calibration. Good agreement was achieved between SALT and X-Shooter fluxes (Table~2), though only the strongest lines are detected in the SALT spectrum.

\begin{table}
\caption{Emission line fluxes in the SALT spectrum.}
\label{lines2}
\centering
{\tiny
\begin{tabular}{c c c c c }
\hline\hline
$\rm \lambda_{obs}$ [\AA]& ident & SALT flux & flux error & X-Shooter flux\\
\hline
4342.570&H\,{\sc i}	&59.4	&	9.3 &	49.67	\\
4365.224&[O\,{\sc iii}]	&25.8	&	7.8 &	33.82	\\
4863.669&H\,{\sc i}	&100.0	&	9.1 &100.00	\\
4961.441&[O\,{\sc iii}]	&86.8	&	8.9 &79.71	\\
5009.392&[O\,{\sc iii}]	&250.1	&	14.8 &237.39	\\
5878.494&He\,{\sc i}	&24.4	&	4.1 &27.73	\\
6303.668&[O\,{\sc i}]	&19.0	&	4.1 &4.25	\\
6566.161&H\,{\sc i}	&442.3	&	22.3 &464.46	\\
6586.981&[N\,{\sc ii}]	&10.7	&	2.5 &9.06	\\
7068.701&He\,{\sc i}	&17.4	&	2.7 &18.67	\\
7139.464&[Ar\,{\sc iii}]&7.3	&	2.6 &3.58	\\
\hline
\end{tabular}
}
\end{table}

[JD2002]\,11 was detected by the Two Micron All Sky Survey (2MASS, Skrutskie et
al., 2006), InfraRed Survey Facility (IRSF, Kato et al., 2007) Wide-Field
Infrared Survey Explorer (WISE, Wright et al., 2010), Deep Near-Infrared Survey
of the Southern Sky (DENIS, Epchtein et al., 1999), and Surveying the Agents of
a Galaxy's Evolution (SAGE-SMC) survey (Gordon et al., 2011) surveys (Table~3).
The SMC was also covered by the Optical Gravitational Lensing Experiment (OGLE)
survey (Udalski et al., 2008), which provides the lightcurve in the I and V
bands covering the period from 1997 to 2014.

\begin{table}
\caption{Optical and infrared photometry of [JD2002]\,11.}
\label{sedtab}
\centering
{\tiny
\begin{tabular}{c c c c c}
\hline\hline
instrument/band	&wavelength [$\rm \mu m$]& magnitude & uncertainty\\
\hline
OGLE V	&	0.54	&	21.554	&	0.289	\\
OGLE I	&	0.79	&	19.55	&	0.126	\\
2MASS J	&	1.25	&	16.067	&	0.083	\\
2MASS H	&	1.65	&	14.07	&	0.04	\\
2MASS Ks&	2.17	&	12.513	&	0.027	\\
WISE	&	3.35	&	10.664	&	0.022	\\
IRAC	&	3.6	&	10.465	&	0.013	\\
IRAC	&	4.5	&	9.761	&	0.011	\\
WISE	&	4.6	&	9.705	&	0.02	\\
IRAC	&	5.8	&	9.146	&	0.012	\\
IRAC	&	8	&	8.527	&	0.01	\\
WISE	&	11.56	&	7.998	&	0.019	\\
WISE	&	22.08	&	7.34	&	0.092	\\
MIPS	&	24	&	6.98	&	0.02	\\
\hline
\end{tabular}
}
\end{table}

\section{Results}

   \begin{figure*}
   \centering
   \includegraphics[width=1.0\textwidth]{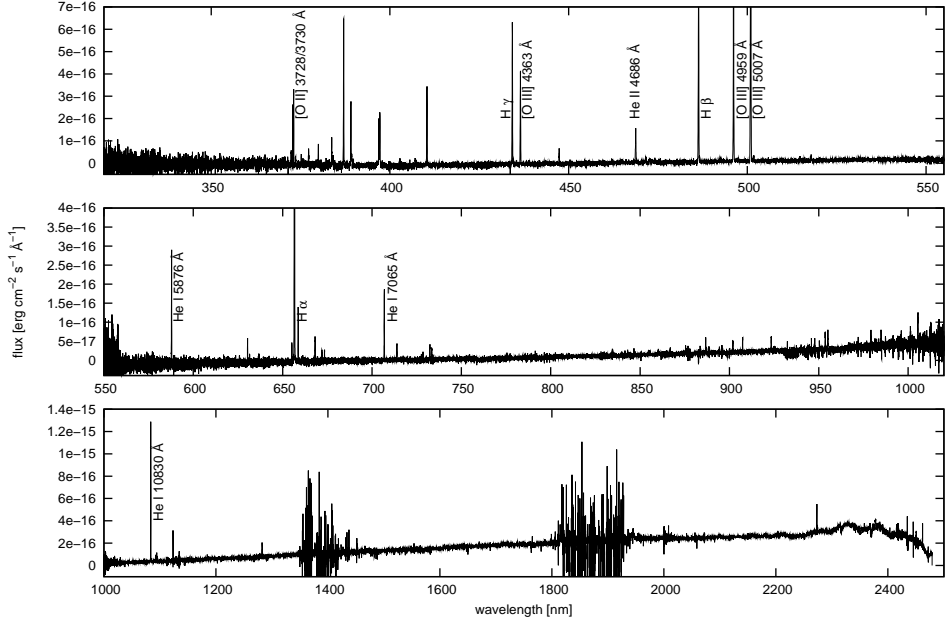}
   \caption{The X-Shooter UVB (top panel), VIS (middle panel) and NIR (bottom panel) spectra of [JD2002]\,11. The residuals for the sky lines and telluric line correction are present in the NIR spectrum.}
              \label{spect}
    \end{figure*}

   \begin{figure*}
   \centering
   \includegraphics[width=0.6\textwidth]{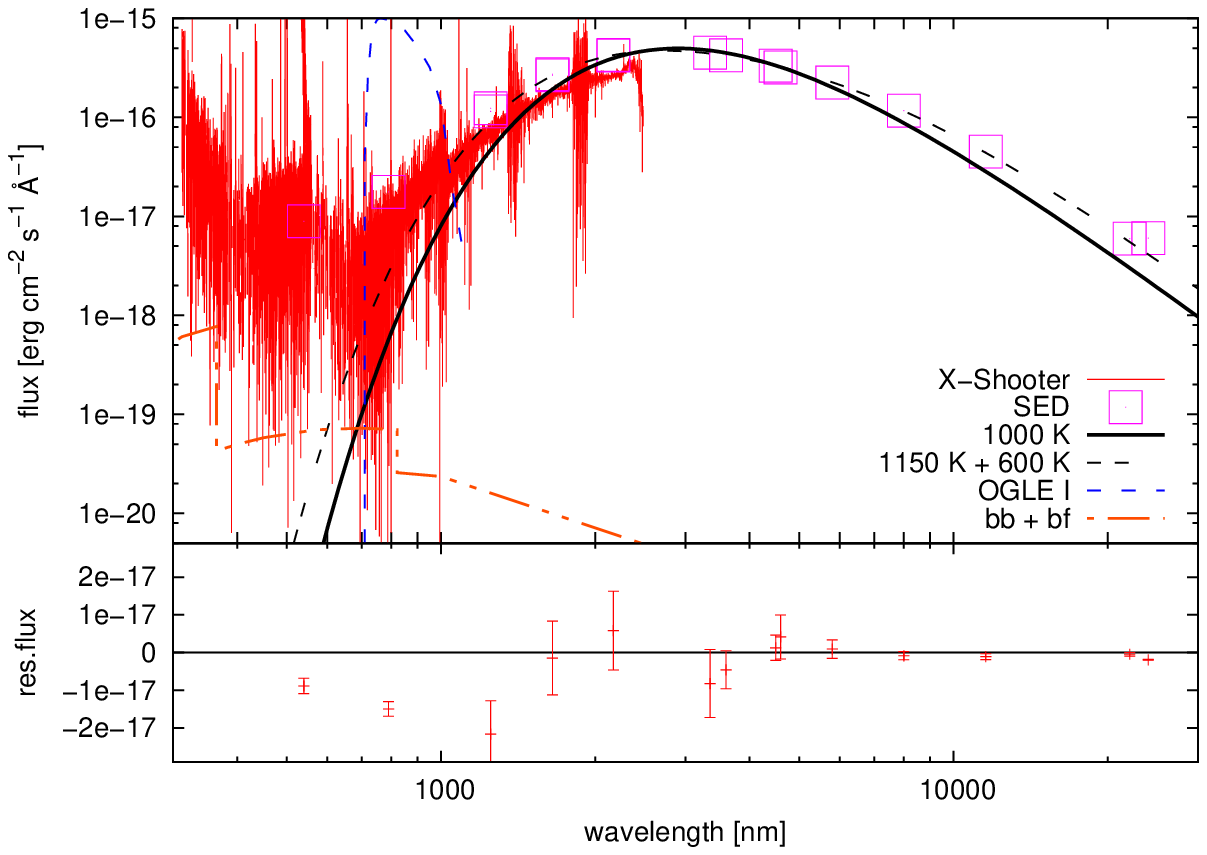}
   \caption{Spectral energy distribution of [JD2002]\,11. Squares show infrared
            photometry from 2MASS, WISE, and Spitzer. The 1000\,K blackbody
            fit is plotted with a thick line. The long-dashed line show the
	    co-added spectrum of two blackbodies with the temperatures of
	    1150\,K and 600\,K. The short-dashed line is the
	    scaled sensitivity of the OGLE I band filter. The dot-dash line
	    shows the contribution of the sum of the bound-bound and bound-free
	    nebular emission. The bottom panel shows the residuals of the two
	    blackbody fit. The error bars correspond to one $\sigma$
	    uncertainties.}
              \label{sed}
    \end{figure*}

[JD2002]\,11 resides in a region of an extended emission bright in the H\,{\sc
i}, [O\,{\sc iii}] 4958/5007\,\AA, [O\,{\sc ii}] 3726/3729\,\AA, and [S\,{\sc
ii}] 6716/6731\,\AA\ lines. The radial velocity of the background emission is
similar to [JD2002]\,11. The background emission most likely originates from the
H\,{\sc ii} (ionized) region DEM S 54. The background emission is smooth along
the slit and was extracted along with the sky emission lines.

The X-Shooter UVB, VIS, and NIR flux calibrated spectra of [JD2002]\,11 are
shown in Figure~1. The spectrum of [JD2002]\,11 shows emission lines
superimposed on a featureless continuum, which rises toward longer wavelengths.
The emission lines are not spatially resolved. Continuum and emission lines
originate from the same position on the slit.

\subsection{Hot component}

The emission lines detected in the spectra of [JD2002]\,11 are listed in
Table~1. We did not detect the O\,{\sc vi} Raman scattered lines at 6825\,\AA\
and 7088\,\AA\ in our spectra (Schmid, 1989). Their presence is restricted
exclusively to symbiotic systems but not all symbiotics show these features.

The observed $\rm H \alpha / H \beta$ line flux ratio is somewhat higher than
the value derived from the Case B recombination. The relative
intensities of $\rm H \alpha$, $\rm H \beta$, and $\rm H \gamma$ emission lines
show, that the main reason for this is self-absorption of the Balmer lines
rather than the circum/interstellar extinction. Thus, for the extinction
correction of the [JD2002]\,11 spectrum we applied the reddening of E(B-V)=0.13
found for the ionized region DEM S 54 (Caplan et al., 1996). We find similar
value of reddening for the $\rm H \alpha / H \beta$ flux ratio measured from the
background emission. The observed and dereddened fluxes of the emission lines
for [JD2002]\,11 are shown in Table~1.

The detection of the He\,{\sc ii} 4686\,\AA\ line indicates the presence of a
relatively hot source in [JD2002]\,11. Its temperature inferred from the
He\,{\sc ii} 4686\,\AA/$\rm H \beta$ line flux ratio is about 100,000\,K. This
is confirmed by the identification of the [Fe\,{\sc vi}] emission line in the
spectrum (Table~1). Ionization of $\rm Fe^{+4}$ requires the photon
energy of 75\,eV.

The [S\,{\sc ii}] 6716/6731\,\AA\ line ratio indicates an electron density of
$\rm 340 \, cm^{-3}$ and the [O\,{\sc ii}] 3726/3729\,\AA\ line flux ratio an
electron density of $\rm 350 \, cm^{-3}$. 

The [O\,{\sc iii}] 4363\,\AA\ emission line is relatively strong compared to the
[O\,{\sc iii}] 5007\,\AA\ line flux. The ratio of $0.153 \pm 0.011$ would
require a spuriously high electron temperature under the electron density {\rm
of about $\rm 350 \, cm^{-3}$}. The same applies to the [S\,{\sc iii}]
6312/9069\,\AA\ line ratio of $0.37 \pm 0.09$. This suggests that the bulk of
the [O\,{\sc iii}] 4363\,\AA\ and [S\,{\sc iii}] 6312\,\AA\ flux originates from
a region of high density of $\rm 10^6 - 10^7\, cm^{-3}$. The [Fe\,{\sc vi}]
5176/5146\,\AA\ line ratio of $1.27 \pm 0.25$ indicates $\rm n_e \sim 10^6 -
10^7\,cm^{-3}$ for $\rm T_e \sim 10-20\,kK$ (e.g. Nussbaumer \& Storey 1978).
He\,{\sc i} and He\,{\sc ii} line flux ratios also indicate $\rm n_e \sim
10^6\,cm^{-3}$. The [S\,{\sc ii}] 6716/6731\,\AA\ and [O\,{\sc ii}]
3726/3729\,\AA\ emission must originate from the more extended region of lower
density. These lines are suppressed in the high density region where their
critical densities are exceeded.

\subsection{Cold component}

The spectral energy distribution (SED) of [JD2002]\,11 is dominated by the
infrared continuum emission (Figure~2). We fitted the datapoints with a single
blackbody function and obtained the temperature of the emitting source of $\rm
1000 \pm 50\,K$ and the corresponding luminosity of $\rm 3000 \pm 200 \,
L_{\odot}$, assuming the distance of $\rm 62.1 \pm 1.9 \, kpc$ to the SMC
derived by Graczyk et al. (2014). The fit reproduced the maximum of the SED
well, but left some residual emission shortward and longward of the maximum
emission. More accurate fit was obtained with the sum of two blackbody functions
with the temperatures of $1150 \pm 50$\,K and $600\pm 100$\,K. The respective
luminosities and radii of the two shells are $\rm 2000 \, L_{\odot}$ and $\rm
1000 \, L_{\odot}$, $\rm 1100 \, R_{\odot}$ and $\rm 2900 \, R_{\odot}$,
assuming that both shells are optically thick.

The two blackbody fit reproduces very well the observed flux between 1.25\,$\rm
\mu m$ and 22\,$\rm \mu m$ (normalized $\chi ^2$ of 0.23). The MIPS 24\,$\rm \mu
m$ data point appears to be overestimated and was not used for the fit.
The SED is dominated by the dust emission. The contribution of the nebular
continuum emission is negligible in the I and V bands. No trace of the
Paschen/Balmer jump is detected in the observed spectra. We did not detect any
molecular absorption bands in the visual and infrared spectra. The X-Shooter
spectra show featureless continuum rising toward red.

Soszy{\'n}ski et al. (2011) obtained a period of 215.2 day and an amplitude of
0.344 in the I band for [JD2002]\,11. The contribution of the variable source
must be at least 30\,\% of the total flux in the I band during the photometric
maximum for the observed amplitude. The contribution of the dust emission fitted
with the two blackbody model to the I band flux is 30\,\% during the maximum.
The contribution of the nebular continuum emission (Figure~2) and emission lines
to the I band flux is negligible.

\subsection{Line profiles}

All spectral features in the X-Shooter spectrum are emissions except for two
helium lines in the NIR region. The He\,{\sc i} line at 10830\,\AA\ shows a
P-Cygni profile, while He\,{\sc i} 20581\,\AA\ line is seen only in absorption
(Figure~4). Both lines originate from the metastable levels of the He\,{\sc i}
atom. The absorption components in the He\,{\sc i} 10830\,\AA\ and 20581\,\AA\
lines are at heliocentric radial velocity of $\rm 100\,km\,s^{-1}$, while the
average radial velocity of the emission lines in [JD2002]\,11 is $\rm
150\,km\,s^{-1}$. The radial velocities of the absorption in the He\,{\sc i}
10830\,\AA\ and 20581\,\AA\ lines are similar to the emission component observed
at the blue wing of the hydrogen Balmer $\rm \alpha$ line (Figure~3).

The dense nebula is optically thick at the He\,{\sc i} 10830\,\AA\ and
20581\,\AA\ lines. However, we would expect  emission lines from both
transitions unless it was formed at the line of sight of the (dust)
thermal radiation emitted by the cold component. So either the absorption lines
are formed in the dense nebula/accretion disk around the hot component, or in
the outer part of the envelope of the cold component which is ionized by the hot
component. In the former case, the hot component must have been placed in the
line of sight to the cold component at the time when the X-Shooter observation
was taken.

   \begin{figure}
   \centering
   \includegraphics[width=0.4\textwidth]{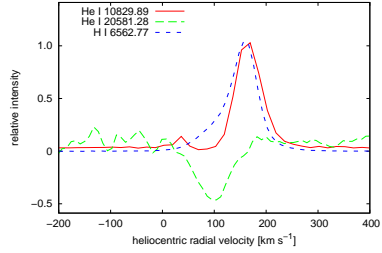}
   \caption{The spectral profiles of two helium lines and the hydrogen line.}
              \label{spect2}
    \end{figure}

\section{Discussion}

The observational characteristics of the cool component confirms that it
is a Mira. The luminosity of the cool component is typical for a Mira.
Almost entire flux of the Mira component is reprocessed by the circumstellar
dust, which is observed in the infrared. The period in the optical corresponds
to the pulsational period of a Mira star. Soszy{\'n}ski et al. (2011)
classified [JD2002]\,11 as a semiregular variable rather than a Mira due to its
low amplitude below 0.8 mag in the I band. However, small amplitude of the
pulsations is typical for symbiotic miras.

The diagnostic line flux ratios of [O\,{\sc iii}] 4363\,\AA/$\rm H\gamma$,
[O\,{\sc iii}] 5007\,\AA/$\rm H\beta$, He\,{\sc i} 6678\,\AA/5876\AA, and
He\,{\sc i} 7065\,\AA/5876\AA\ place [JD2002]\,11 in the region occupied by the
D-type symbiotic stars (Gutierrez-Moreno, 1995; Proga et al., 1994).
[JD2002]\,11 falls into the region occupied by heavily obscured D-type symbiotic
stars in the (J-H) - (H-$\rm K_s$) colour-colour diagram (Feast et al., 1977).
The presence of the two dust component is also typical for D-type
symbiotic stars (Angeloni et al., 2010).

The near-infrared spectrum of [JD2002]\,11 is typical for symbiotic Miras during
high dust obscuration phase, e.g. V835\,Cen (Feast et al. 1983). In particular,
the molecular absorption bands are absent and dust emission dominates the
continuum, which steeply rises toward infrared in both objects.

   \begin{figure}
   \centering
   \includegraphics[width=1.0\textwidth]{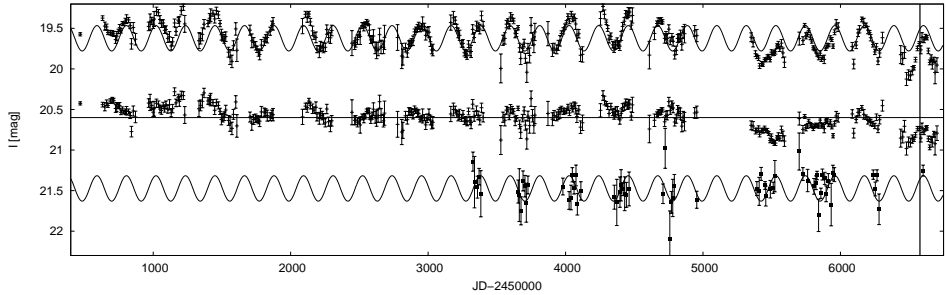}
   \caption{The OGLE lightcurve of [JD2002]\,11 in the I (top) and V band (bottom) with the 215-day period plotted. The residuals after the subtraction the sinusoidal fit from the I band lightcurve are plotted in the middle. The vertical bar marks the date of the X-Shooter observation.}
              \label{lightcurve2}
    \end{figure}

The pulsations in near-IR are present in V835 Cen and another symbiotic
Mira, RX Pup, even during high dust-obscured phase while they are invisible in
the optical (Feast et al., 1983; Mikolajewska et al., 1999). RX Pup is
permanently embedded in an optically thick dusty shell and never shows evidence
for a cool star shortwards of $\rm \sim 800 \, nm$, and it also never shows
pulsations in this range (Mikolajewska et al. 1999). This is because the
spectrum shortwards of $\rm \sim 800\,nm$ is dominated by a nebular emission. We
do not observe the spectral signature of the cool star in [JD2002]\,11 as well.
The lack of pulsations in the V band may be caused by the contribution of the
nebular lines and continuum. Since we do not observe any significant change
(exceeding 0.1 mag) in the near-IR fluxes between 2MASS, IRSF and DENIS
observations, we conclude that [JD2002]\,11 is permanently obscured by dust.

[JD2002]\,11 shows similar characteristics to a group of objects discussed by
Rodr\'{\i}guez-Flores et al. (2014), which have been classified as D-type
symbiotic stars. They show no direct evidence of the cold component in their
optical spectra. Miszalski et al. (2013) identified some new D-type symbiotic
stars for which no signature of cold component was found in the spectroscopy,
but which show photometric variability.

P Cygni profile of the He\,{\sc i} 10830\,\AA\ line are reported for two out of
eight S-type symbiotic stars by Baratta et al. (1991). They attributed it to the
variable wind from the hot component. They also observed He\,{\sc ii}
10123\,\AA\ emission line for only two objects: high excitation symbiotic stars
Z\,And and AG\,Peg.

The lightcurve of [JD2002]\,11 reveals an irregular variability in addition to
the 215 day Mira pulsation period, which is shown in Figure~4. We
attribute this to the variability of Mira star since it dominates in this
spectral region. 

The period of 215 days is shorter than in any of the Galactic D-type
symbiotic systems (Gromadzki et al., 2009). It is shorter than the shorthest
pulsation period in the grid of evolutionary models for (single) AGB stars
computed by Vassiliadis and Wood (1993) for the SMC (Z=0.004). However, LMC
D-type systems show even shorter periods (Angeloni et al., 2014). The
bolometric magnitude obtained for [JD2002]\,11 of $-3.91$ is slightly lower than
the maximum luminosity for thermally pulsating AGB star of $-3.98$ obtained by
Vassiliadis and Wood (1993) for a star with a maximum period of 310 days. The
period and luminosity of the giant in [JD2002]\,11 suggest an initial mass of
about $\rm 0.9\,M_{\odot}$, assuming that it evolved accordingly to single AGB
evolutionary tracks. [JD2002]\,11 obeys the period-luminosity relation derived
by Groenewegen and Whitelock (1996).

We conclude that [JD2002] 11 is the ninth symbiotic star in total and only
the second dusty type symbiotic star discovered in the Small Magellanic Cloud. 
it is located in the SMC bar. Interestingly, the other D-type symbiotic star is
also placed in the SMC bar. However, the sample is too small to draw firm
conclusions.

\Acknow{
This work was financially supported by NCN of Poland through grants No.
2011/01/D/ST9/05966 and 719/N-SALT/2010/0. Some of the observations reported in
this paper were obtained with the Southern African Large Telescope (SALT). MG
acknowledges support from Joined Committee ESO and Government of Chile 2014 and
the Ministry for the Economy, Development, and Tourisms Programa Inicativa
Cientifica Milenio through grant IC 12009, awarded to The Millennium Institute
of Astrophysics (MAS) and Fondecyt Regular No. 1120601. This publication makes
use of data products from the Two Micron All Sky Survey, which is a joint
project of the University of Massachusetts and the Infrared Processing and
Analysis Center/California Institute of Technology, funded by the National
Aeronautics and Space Administration and the National Science Foundation. This
publication makes use of data products from the Wide-field Infrared Survey
Explorer, which is a joint project of the University of California, Los Angeles,
and the Jet Propulsion Laboratory/California Institute of Technology, funded by
the National Aeronautics and Space Administration. his work is based in part on
observations made with the Spitzer Space Telescope, which is operated by the Jet
Propulsion Laboratory, California Institute of Technology under a contract with
NASA. This publication makes use of VOSA, developed under the Spanish Virtual Observatory project supported from the Spanish MICINN through grant AyA2011-24052.
}

\end{document}